\documentclass[prc,nofootinbib,superscriptaddress,twocolumn]{revtex4}
\usepackage{graphicx,dcolumn,array,bm}

\newcommand{\tfrac}[2]{{\textstyle{\frac{#1}{#2}}}}
\newcommand{\svec}[1]{\bm{#1}}
\newcommand{\ivec}[1]{\vec{#1}}

\begin{document}
\bibliographystyle{apsrev}


\title{Time-reversal violating Schiff moment of $^{225}$Ra}
\author{J. Engel}
\email[]{engelj@physics.unc.edu}
\affiliation{Department of Physics and Astronomy, CB3255, 
             University of North Carolina, Chapel Hill, NC  27599-3255}
\author{M. Bender}
\email[]{mbender@ulb.ac.be}
\affiliation{Service de Physique Nucl{\'e}aire Th{\'e}orique 
            et de Physique Math{\'e}matique, 
            Universit{\'e} Libre de Bruxelles -- C.P. 229, 
            B-1050 Brussels, Belgium}
\author{J. Dobaczewski}
\email[]{Jacek.Dobaczewski@fuw.edu.pl}
\affiliation{Institute of Theoretical Physics, 
             Warsaw University, 
             Ho\.za 69, PL-00681, Warsaw, Poland}
\author{J.H. de Jesus}
\email[]{jhjesus@physics.unc.edu}
\affiliation{Department of Physics and Astronomy, CB3255, 
             University of North Carolina, Chapel Hill, NC  27599-3255}
\author{P. Olbratowski}
\email[]{Przemyslaw.Olbratowski@fuw.edu.pl}
\affiliation{Institute of Theoretical Physics, 
             Warsaw University, 
             Ho\.za 69, PL-00681, Warsaw, Poland}
\affiliation{Institut de Recherches Subatomiques,
             UMR7500, CNRS-IN2P3 and Universit\'e Louis Pasteur,
             F-67037 Strasbourg Cedex 2, France}
\date{\today}
\begin{abstract}
We use the Skyrme-Hartree-Fock method, allowing all symmetries to be
broken, to calculate the time-reversal-violating nuclear Schiff moment
(which induces atomic electric dipole moments) in the octupole-deformed
nucleus $^{225}$Ra. Our calculation includes several effects neglected in
earlier work, including self consistency and polarization of the core by
the last nucleon.  We confirm that the Schiff moment is large compared to
those of reflection-symmetric nuclei, though ours is generally a few times
smaller than recent estimates.
\end{abstract}
%
%
\maketitle
%
%
\section{Introduction}
\label{s:intro} 
Experiments with K and B mesons indicate that time-reversal invariance (T) is
violated through phases in the Cabibbo-Kobayashi-Maskawa matrix that affect
weak interactions \cite{kowalewski03}.  The suspicion that
extra-Standard-Model
physics, e.g.\ supersymmetry, also violates T has motivated a different kind
of experiment:  measuring the electric dipole moments (EDMs) of the neutron
and of atoms.  Because any such dipole moment must be proportional to the
expectation value of the T-odd spin operator, it can only exist when T (and
parity) is violated \cite{sachs87,khriplovich97}.  So far the experiments have
seen no dipole moments, but they continue to improve and even null results are
useful, since they seriously constrain new physics.  Whatever the experimental
situation in the future, therefore, it is important to determine theoretically
what the presence or absence of EDMs at a given level implies about
T-violating interactions at elementary-particle scales.  Our focus here is
atoms, which for some sources of T violation currently provide limits as good
or better than the neutron \cite{romalis01}.

One way an atom can develop an EDM is through T and P violation in its
nucleus.  Let us assume that given a fundamental source of the broken symmetry
one can use effective-field theory and QCD to calculate the strength of the
resulting T-violating nucleon-pion interaction.  One then needs to connect the
strength of that interaction to the resulting nuclear ``Schiff moment'',
which, because the nuclear EDM is screened \cite{schiff63}, is the quantity
responsible for inducing an EDM in electrons orbiting the nucleus.  The Schiff
moment is defined classically as a kind of radially weighted dipole moment:
\begin{equation}
\label{eq:mom}
\svec{S}
= \tfrac{1}{10} \int {\rm d}^3r \rho_{\rm ch}(\svec{r}) 
  \left(r^2 - \tfrac{5}{3} \overline{r_{\rm ch}^2}  \right) \svec{r}
,
\end{equation}
where $\rho_{\rm ch}$ is the nuclear charge density and
$\overline{r_{\rm ch}^2} $ is the mean-square charge radius.  Recent papers
\cite{spevak97,auerbach96} have argued that because of their asymmetric
shapes, octupole-deformed nuclei in the light-actinide region should have
collective Schiff moments that are 100 to 1000 times larger than the Schiff
moment in $^{199}$Hg, the system with the best experimental limit on its
atomic EDM \cite{romalis01}.  Ref.\ \cite{engel99a} suggested that certain
many-body effects may make the enhancement a bit less than that.  The degree
of enhancement is important because several experiments in the light 
actinides
are contemplated, planned, or underway \cite{chupp01,holt03}.  They may see
nonzero EDMs, and even if they don't we will need to be able to compare their
limits on fundamental sources of T violation to those of experiments in other
isotopes.

Perhaps the most attractive octupole-deformed nucleus for an experiment is
$^{225}$Ra.  Though radioactive, it has a ground-state angular momentum
$J=1/2$, which minimizes the effect of stray quadrupole electric fields in an
experiment to measure a dipole moment\footnote{The statement that the nucleus
has octupole and quadrupole deformation really refers to its
\textit{intrinsic} state, a concept we elaborate on below, and does not
contradict its insensitivity to applied electric fields with multipolarity
greater than one.}.  In addition, the associated atom has close-lying
electronic levels of opposite parity and is relatively easy to trap and
manipulate.  As a result, at least one group is at work on a measurement in
$^{225}$Ra \cite{holt03}.  Here we calculate its Schiff moment, attempting to
incorporate the effects discussed in Ref.\ \cite{engel99a} through a
symmetry-unrestricted mean-field calculation.  We begin in the next section 
by
describing the physics of the Schiff moment in octupole-deformed nuclei,
briefly reviewing prior work in the process.  In Section \ref{s:3} we test 
our
mean-field approach by calculating properties of even Ra isotopes.  In 
Section
\ref{s:4} we discuss issues peculiar to mean-field calculations in odd nuclei
and then present our results for the Schiff moment of $^{225}$Ra, focusing
particularly on the degree of enhancement.  Section \ref{s:5} is
a brief conclusion.
%
%
\section{\label{s:2}Enhancement of Schiff Moments in Octupole-Deformed Nuclei
                    -- Previous Work}
In analogy with dipole moments in atoms, static Schiff moments in nuclei can 
exist only if T is broken.  Because T-violating forces are much weaker than 
the 
strong interaction, the Schiff moment can be accurately evaluated through 
first-order perturbation theory as
\begin{equation}
\label{eq:def} 
S \equiv \langle \Psi_0 | \hat{S}_z | \Psi_0 \rangle = \sum_{i \neq 0} 
         \frac{\langle \Psi_0 | \hat{S}_z |\Psi_i \rangle 
               \langle \Psi_i | \hat{V}_{PT} | \Psi_0 \rangle} 
              {E_0 - E_i} 
         + \text{c.c.} 
,
\end{equation}
where $| \Psi_0 \rangle$ is the member of the ground-state multiplet with 
$J_z=J\neq0$, the sum is over excited states, and $\hat{S}_z$ is the operator
\begin{equation}
\hat{S}_z 
= \tfrac{e}{10} \sum_p \left( r_p^2 - 
    \tfrac{5}{3} \overline{r_{\rm ch}^2}  \right) \, z_p
,
\end{equation}
with the sum here over protons. The operator $\hat{V}_{PT}$ is the T- (and
parity-) violating nucleon-nucleon interaction mediated by the pion
\cite{haxton83,herczeg88} (shown to be more important than other mesons in
Ref.\ \cite{towner94}):
\begin{widetext}
\begin{eqnarray}
\label{eq:pion} 
\hat{V}_{PT}(\svec{r}_1-\svec{r}_2) 
               & = - {\displaystyle   \frac{g \, m_{\pi}^2}{8 \pi m_N}} &
                     \Big\{
                         (\svec{\sigma}_1 - \svec{\sigma}_2) \cdot
                         (\svec{r}_1-\svec{r}_2) 
                         \left[ \bar{g}_0 \, \ivec{\tau}_1 \cdot
                                             \ivec{\tau}_2 
                         - \frac{\bar{g}_1}{2} \,
                          (\tau_{1z}+\tau_{2z}) + \bar{g}_2
                         (3\tau_{1z}\tau_{2z} - \ivec{\tau}_1 \cdot
                         \ivec{\tau}_2 ) \right]
                          \nonumber   \\
                        && -\frac{\bar{g}_1}{2}
                         (\svec{\sigma}_1+\svec{\sigma}_2)
                         \cdot (\svec{r}_1-\svec{r}_2) \,
                         (\tau_{1z}-\tau_{2z})  \Big\}
\frac{{\rm exp}(-m_{\pi} |\svec{r}_1-\svec{r}_2|)}{m_{\pi}
|\svec{r}_1-\svec{r}_2|^2} \left[ 1+\frac{1}{m_{\pi}|\svec{r}_1-\svec{r}_2|}
\right] 
,
\end{eqnarray}
\end{widetext}
where arrows denote isovector operators, $\tau_z$ is +1 for neutrons,
$m_N$ is the nucleon mass, and we 
are using 
the convention \mbox{$\hbar = c = 1$}.  The $\bar{g}$'s are the unknown 
isoscalar, isovector, and isotensor T-violating pion-nucleon
couplings, and $g$ is the usual strong ${\pi NN}$ coupling.

In a nucleus such as $^{199}$Hg, with no intrinsic octupole deformation, many 
intermediate states contribute to the sum in Eq.\ (\ref{eq:def}).   By 
contrast, the asymmetric shape of  $^{225}$Ra implies the existence of a very 
low-energy $|1/2^-\rangle$ state, in this case 55 keV above the ground state 
$|\Psi_0\rangle \equiv |1/2^+\rangle$, that dominates the sum because of the 
corresponding small denominator. To very good approximation, 
then,
\begin{equation}
\label{eq:approx} 
S \equiv - \frac {\langle 1/2^+|\hat{S}_z|1/2^- \rangle
         \langle 1/2^- | \hat{V}_{PT} | 1/2^+ \rangle} {\Delta E} + 
\text{c.c.} 
,
\end{equation}
where $\Delta E$ = 55 keV.  The small denominator is part of the reason for 
the 
enhancement of the Schiff moment.  The other part is the matrix element of 
the 
Schiff operator in Eq.\ (\ref{eq:approx}).  In the limit that the deformation   
is 
rigid, the 
ground state and first excited state in octupole-deformed nuclei are  
partners 
in a parity doublet, \textit{i.e.}, projections onto good parity and angular 
momentum of the same ``intrinsic state" that represents the wave function of 
the nucleus in its own body-fixed frame.  The matrix elements in Eq.\ 
(\ref{eq:approx}) are then proportional (again, in the limit of rigid 
deformation) to intrinsic-state expectation values, so that \cite{spevak97}
\begin{equation}
\label{eq:intr} S \longrightarrow - 2 \frac{J}{J+1} \frac {\langle
\hat{S}_z \rangle
\langle  \hat{V}_{PT}  \rangle} {\Delta E}  \ \ ,
\end{equation}
where $J$ is the ground-state angular momentum, equal to 1/2 for $^{225}$Ra, 
and
the brackets indicate expectation values in the intrinsic state.  The
intrinsic-state expectation value $\langle \hat{S}_z \rangle$ is generated by
the collective quadrupole and octupole deformation of the entire nucleus; it
is much larger than a typical matrix element in a spherical or
symmetrically deformed nucleus.  Together with the small energy denominator,
this large matrix element is responsible for the enhancement of 
laboratory-frame
Schiff moments in nuclei such as $^{225}$Ra.

The amount of the enhancement is not easy to calculate accurately, however. 
The 
reason is that the matrix element of the two-body spin-dependent operator
$\hat{V}_{PT}$ in Eq.\ (\ref{eq:approx}) depends sensitively on the behavior 
of a few valence particles, 
which carry most of the spin.  In the approximation that particles (or
quasiparticles) move in independent orbits generated by a mean field, the 
potential can be written as an effective density-dependent one-body operator 
that we will denote $\hat{U}_{PT}$, defined implicitly by
\begin{equation}
\label{eq:spdef}
\langle a|\hat{U}_{PT}|b\rangle 
= \sum_{c<F} \langle a c|\hat{V}_{PT} | b c \rangle
,
\end{equation}
where $|a\rangle$, $|b\rangle$, and $|c\rangle$ are eigenstates of the mean
field and the matrix elements of $\hat{V}_{PT}$ are antisymmetrized.  With
the further approximation that the mass of the pion is very large, 
$\hat{U}_{PT}$ 
can be written as a local operator, in a form we display in the Section 
\ref{s:4}.  Evaluating its matrix element is tricky.

The authors of Refs.\ \cite{spevak97,auerbach96} used a version of the 
particle-rotor model \cite{leander84} to represent the odd-$A$ nucleus.  In 
this 
model, all but one of the nucleons are treated as a rigid core, and the last 
valence nucleon occupies a deformed single-particle orbit, obtained by 
solving 
a Schr\"odinger equation for a Nilsson or deformed Wood-Saxon potential.  The 
model implies that the core carries no intrinsic spin whatever, that the 
neutron and proton densities are proportional, and that the exchange terms on 
the right-hand side of Eq.\ (\ref{eq:spdef}) are negligible.  Under these 
assumptions, $\hat{U}_{PT}$, which now acts only on the single valence 
nucleon, reduces to \cite{sushkov84}
\begin{equation}
\label{eq:sppot} 
\hat{U}_{PT}(\svec{r}) 
\approx \eta \frac{G}{2 m_N \sqrt{2}}
        \svec{\sigma} \cdot \svec{\nabla} \rho_0(\svec{r})
.
\end{equation}
where $G$ is the Fermi constant,
inserted to follow convention, and $\rho_0$ is the total nuclear mass 
density.  The dimensionless parameter $\eta$ is
then a function of the couplings $\bar{g}_i$ and the isospin of the nucleus.

Ref.\ \cite{engel99a} confirmed the collectivity of the intrinsic Schiff 
moments
obtained in Refs.\ \cite{spevak97,auerbach96}, but questioned the accuracy of 
some of the
assumptions used to evaluate the matrix element of $\hat{V}_{PT}$, suggesting
that either core-spin polarization or self-consistency in the
nuclear wave function might reduce laboratory Schiff moments. The zero-range
approximation and the neglect of exchange in $\hat{U}_{PT}$ are also open
to question.  As a result, it is not clear whether the Schiff moment of 
$^{225}$Ra is 1000 times that of $^{199}$Hg or 100 times, or even less.  In 
what 
follows, we provide a (tentative) answer by moving
beyond the particle-rotor model. Our calculation is not the final word on
Schiff moments in octupole-deformed nuclei --- we only do mean-field
theory, neglecting in particular to project onto states with good parity, and 
do not fully account for the pion's nonzero range --- but is a major step 
forward.
%
%
\section{Mean-field calculations for other R\lowercase{a} isotopes}
\label{s:3}
\begin{figure*}[thp]
\includegraphics{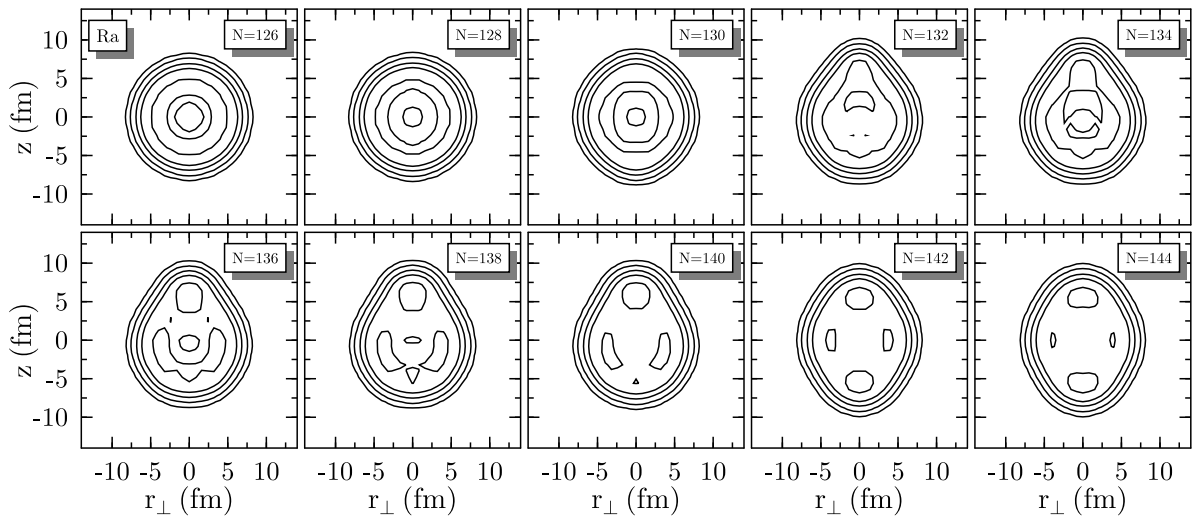}
\caption{\label{fig:dens} Contours of constant density for a series
of even-$N$ Radium isotopes.  Contour lines are drawn for densities
$\rho$=0.01, 0.03, 0.07, 0.11, and 0.15 fm$^{-3}$.}
\end{figure*}
%
%
\subsection{Mean-field calculations}
Self-consistent mean-field theory is widely used for describing bulk 
properties of nuclei \cite{bender03}.  In the guise of density-functional 
theory, it is also used throughout atomic and molecular physics. 
The approach is more ``microscopic" --- nucleons are the only degrees of 
freedom --- and far less phenomenological
than the collective particle-rotor model. Self-consistency connects 
the single-particle states and the actual density distribution. The 
variational
principle that determines the single-particle wave functions thus optimizes  
all multipole moments not fixed by global symmetries. 
The density distributions of neutrons and protons are not proportional 
to each other; they have slightly different deformations and radial profiles. 
In odd-$A$ nuclei, self-consistent calculations include rearrangement due 
to the unpaired particle.  Rearrangement causes polarization of the 
even-$A$ core through orbital-current and spin-density terms 
in the effective interaction.  Core polarization is one of the effects on the 
Schiff moment of $^{225}$Ra that we investigate below.

Our approach is nonrelativistic and employs Skyrme interactions. 
To get an idea of the range of results this kind of calculation 
can produce, we use four different parameterizations of the Skyrme energy 
functional, i.e., four different Skyrme forces.  The four give similar 
results 
for many observables near stability, 
but still have significant differences. Our favorite interaction, for reasons 
explained below, is SkO' \cite{bender02,reinhard99}, but we also show results 
for the commonly 
used forces SIII \cite{beiner75}, SkM$^*$ \cite{bartel82}, and 
SLy4 \cite{chabanat98}.

%
%
\subsection{Related Observables in Even Isotopes}
Intrinsic-parity breaking in even radium isotopes is the subject of several
theoretical analyses; see the review in Ref.\ \cite{butler96} and the more
recent studies in Refs.\ \cite{garrote97,tsvetkov02}.  To assess the ability
of the Skyrme interactions to handle it, we perform a series of Hartree-Fock
(HF) + BCS calculations for even Radium isotopes.  We use the Skyrme-HF+BCS
code from Ref.\ \cite{bender98}; it represents single-particle wave functions
on an axially symmetric mesh, and uses Fourier definitions of the derivative,
$1/r_\perp$, and $1/r_\perp^2$ operators.  We choose 75 grid points in the $z$
direction, and 27 in the $r_\perp$ (perpendicular) direction, with 0.8~fm
between them.  The code uses a density-independent zero-range pairing
interaction with a self-adjusting cutoff as described in Ref.\ 
\cite{bender00}.
For each Skyrme force we adjust the pairing strength separately for protons
and neutrons \cite{bender00}.  We should note that other self-consistent
mean-field models, namely HF+BCS with the nonrelativistic Gogny force
\cite{egido89} and the relativistic mean-field model \cite{rutz95}, yield
results that are similar to those we describe now.

Figure \ref{fig:dens} illustrates the calculated evolution of intrinsic
deformation with increasing neutron number in the Radium isotopes.  It plots
the intrinsic ground-state mass-density contours predicted by SkO'.  The
mean-field ground states go from having a spherical shape at the magic number
\mbox{$N=126$} to a quadrupole deformed (reflection-symmetric) shape at
\mbox{$N=130$}, then to quadrupole+octupole deformed (reflection-asymmetric)
shapes for \mbox{$132 \leq N \leq 140$}, and finally back to quadrupole
deformed shapes at higher $N$.  Because the ground states are obtained from a
variational principle, all shape moments higher than octupole are also
optimized (the isoscalar dipole moment is constrained to be zero).  The
nucleus $^{225}$Ra, with \mbox{$N=137$}, will clearly be well deformed in 
both
the quadrupole and octupole coordinates.  The structures at small radii
visible for \mbox{$N \geq 132$} reflect small oscillations of the density
distribution around the saturation value (for a given neutron excess) caused
by shell effects.
\begin{figure}[htp]
\includegraphics{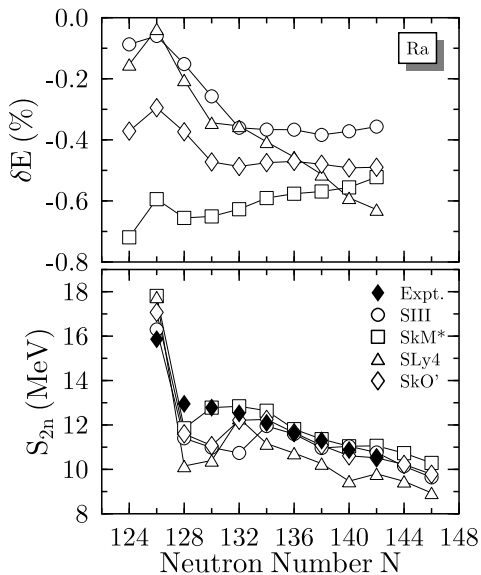}
\caption{\label{fig:s2n}Relative error in binding energy (top) and predicted
two-neutron separation energies (bottom) for four Skyrme interactions in a
series of even-$N$ Radium isotopes.  The experimental separation energies are
also shown.}
\end{figure}

We must note that the octupole-deformed minima are not equally pronounced for
all forces and isotopes.  In addition, in some of the isotopes with
reflection-symmetric minima, some of the Skyrme forces predict an excited
octupole-deformed minimum separated by a small barrier from the ground-state
minimum.  Furthermore, in the transitional nuclei, which have soft potentials
in the octupole direction, all parity-breaking intrinsic deformations are
subject to collective correlations as discussed in Ref.\ \cite{egido89}.  The
influence of correlations will be smallest for the nuclides with the most
pronounced octupole-deformed minima, usually $^{222}$Ra and $^{224}$Ra.  This
fact supports our belief that our mean-field calculations supply a good
approximation to the intrinsic structure in $^{225}$Ra.

Figure \ref{fig:s2n} shows the relative error in the predicted binding 
energies
$\delta E = (E_{\text{calc}}-E_{\text{expt}})/E_{\text{expt}}$
for all four forces, and the predicted two-neutron separation energies,
along with the measured values. All the forces do a good job with binding,
which is not surprising given the way their parameters were fit. The fact
that the error in binding for SkO' is nearly constant with $N$ for $N > 130$ 
is reflected
in the near perfect agreement in the bottom panel with the measured
two-neutron separation energies $S_{2n}$.
The errors in predicted values of $S_{2n}$ around \mbox{$N=128$} 
probably reflect the deficiencies of mean-field models in transitional 
nuclei.
\begin{figure}[htp]
\includegraphics{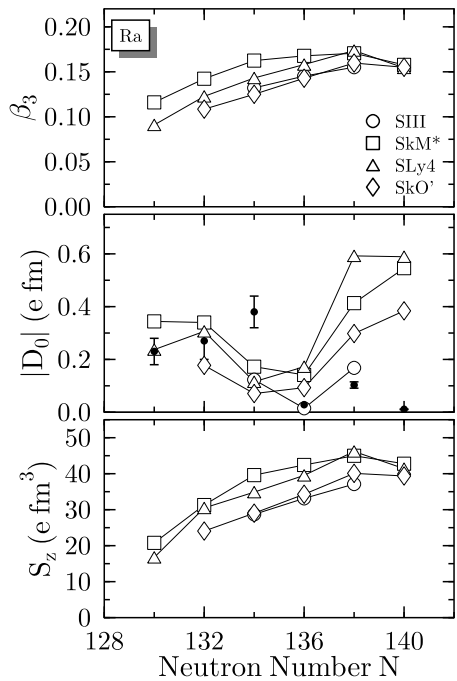}
\caption{\label{fig:def} The predicted first-order \cite{ring80} octupole
deformations (top), intrinsic dipole moments (middle) and intrinsic Schiff
moments (bottom) for four Skyrme interactions in a series of even-$N$ Radium
isotopes.  The experimental intrinsic dipole moments are also shown.  Where
symbols are missing, the corresponding predicted values are zero because the
mean field is not asymmetrically deformed.}
\end{figure}

Figure \ref{fig:def} shows three parity-violating intrinsic quantities.  In
the top panel is the ground-state octupole deformation $\beta_3 = 4 \pi
\langle r^3 Y_{30} \rangle / (3 A R^3)$ (where \mbox{$R=1.2 A^{1/3}$}), as a
function of neutron number.  The trend mirrors that in the density profiles
shown earlier.  At \mbox{$N=136$}, one less neutron than in $^{225}$Ra, the
forces all predict almost identical octupole deformation, a result we like.
Experimental data for octupole moments are still sparse in this region; we 
are
only aware of \mbox{$\beta_3 = 0.105 (4)$} for \mbox{$N=138$}, a value that
can be deduced from the $B(E3;0^+_1 \to 3^-_1)$ given in Ref.\
\cite{spear89}.  (In $^{224}$Ra and $^{226}$Ra, by the way, we agree fairly
well with the quadrupole moments obtained from $B(E2)$'s in Ref.\
\cite{raman01}.  For example, SkO' gives \mbox{$\beta_2 = 0.184$} in
$^{224}$Ra and experiment gives \mbox{$\beta_2 = 0.179(4)$}.)

The second panel in the figure shows the absolute values of intrinsic dipole
moments \mbox{$D_0 = e \sum_p \langle z_p \rangle$}, along with experimental
data extracted from $E1$ transition probabilities \cite{butler96}.  The
calculated values for $D_0$ change sign from positive to negative between
\mbox{$N=134$} and \mbox{$N=138$}, reflecting a small change in the location 
of the center of charge from the ``top" half of the pear-shaped nucleus to the 
``bottom" half.  This predicted
sign change is consistent with the near-zero experimental value for
\mbox{$N=136$}.  None of the forces precisely reproduces the trend through 
all
the isotopes, but the comparison has to be taken with a grain of salt because
``data" derive from transitions between excited rotational states, and
therefore are not necessarily identical to the ground-state dipole moments.
Cranked Skyrme-HF calculations without pairing correlations \cite{tsvetkov02}
and cranked HFB calculations with the Gogny force \cite{garrote97} predict
that for most Ra isotopes $D_0$ changes significantly with angular momentum.
In any event, as thoroughly discussed in Ref.\ \cite{butler96}, the intrinsic
dipole moment is a small and delicate quantity.

The intrinsic Schiff moment \mbox{$\langle S_z \rangle$}, the quantity we're
really interested in, is more collective and under better control, as the
bottom panel of the figure shows.  The various predictions are usually within
20$\%$ of one another and large, confirming the predictions originally made in
Refs.\ \cite{spevak97,auerbach96}.  The octupole deformation and intrinsic
dipole moment
have been shown to change only slightly with parity projection from the
intrinsic states \cite{garrote97}, and the same is probably true of the
intrinsic Schiff moment.

By turning the pairing force off, we are able to see whether the
parity-violating quantities in Fig.\ \ref{fig:def} are affected by pairing 
correlations. In $^{224}$Ra, for example, SkO' gives
\mbox{$\beta_3=0.141$}, \mbox{$D_0=-0.103$}~$e$\,fm, and 
\mbox{$\langle S_z \rangle =34.4$}~$e$\,fm$^3$ without pairing, and
\mbox{$\beta_3=0.143$}, \mbox{$D_0=-0.093$}~$e$\,fm, and
\mbox{$\langle S_z \rangle =34.3$}~$e$\,fm$^3$ when pairing is included. In 
this nucleus 
uncertainties related to pairing are very small.
\begin{figure}[htp]
\includegraphics{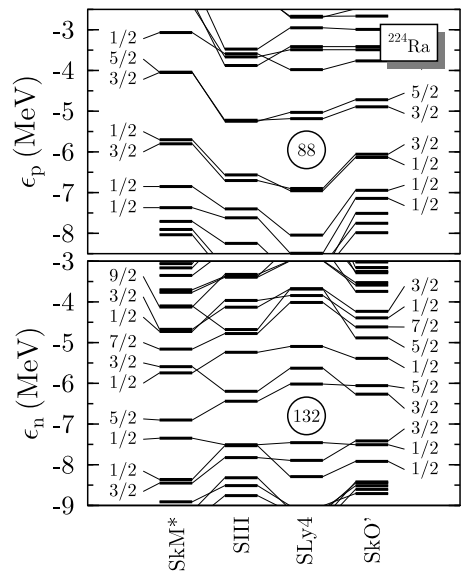}
\caption{\label{fig:sp}
Single-particle spectra for protons (top) and neutrons
(bottom) in $^{224}$Ra, for the four Skyrme interactions.}
\end{figure}

Finally, in Fig.\ \ref{fig:sp}, we show the predicted proton and neutron
single-particle spectra generated by the ground-state mean-field in
$^{224}$Ra.  The combination of quadrupole, octupole, and higher deformations
reduces the level density around the Fermi surface for both kinds of nucleon,
leading to significant deformed \mbox{$Z=88$} and \mbox{$N=132$} shell
closures for all interactions, and a somewhat weaker \mbox{$N=136$} subshell
closure for SIII, SkM* and SkO'.  The small level density around the Fermi
surface might explain the insensitivity of the deformation to pairing
correlations mentioned above.  For all the forces except SkM$^*$, the first
empty neutron level clearly has \mbox{$j_z=1/2$}, implying that in $^{225}$Ra
the ground-state parity-doublet bands will be built on \mbox{$J^\pi=1/2^\pm$}
states.  For SkM$^*$ the situation is less clear because the \mbox{$j_z=1/2$}
and 3/2 states are nearly degenerate, and it is necessary to carry out the
calculation in $^{225}$Ra itself to see which becomes the lowest.
%
%
\section{\label{s:4}Calculating the Schiff moment of $^{225}$R\lowercase{a}}
%
%
\subsection{Odd-$A$ Nuclei and Schiff Moments in Mean-Field Approximation}
Fully self-consistent calculations in odd-$A$ nuclei are possible
but seldom performed. For many physical observables it is enough to
neglect correlations between the odd particle and the core,
which amounts to dropping a valence particle into the field generated
by that core. Other quantities, however, are sensitive to 
the interaction between the last particle and the core.  The interaction 
can change the deformation and pairing strength, and produce various kinds of 
core polarization.  
In fact, a self-consistent odd-$A$ calculation is equivalent to first 
performing the calculation in the even-even nucleus with one less neutron, 
then placing the last nucleon in the first empty orbit and treating its
polarizing effect on the core in RPA \cite{brown70,blaizot86}. Fully
self-consistent calculations take all kinds of polarization into account
simultaneously.  For us, the spin polarization is of particular interest.

To proceed here we use the public-domain Skyrme-Hartree-Fock code HFODD
(v2.04b) \cite{dobaczewski97a,dobaczewski00a,olbratowski03}, which allows the
mean-field and the associated Slater determinant to simultaneously break
invariance under rotation, reflection through three planes, and time reversal 
(the code we used in the last section does not).  Breaking the first
is necessary to describe a deformed nucleus.  Breaking all three
reflections (and not only parity) is necessary to represent
axial octupole deformation with the spin aligned along the symmetry
axis. Breaking the last induces spin polarization in the core, which
because of Kramers degeneracy cannot otherwise occur.  
Incorporating spin polarization is important because it has the potential to
significantly alter the matrix element of $\hat{U}_{PT}$ in Eq.\
(\ref{eq:sppot}) from its value in the particle-rotor model, where the spin is
carried entirely by one valence particle.  The code HFODD cannot yet treat
pairing when it allows T to be broken, but pairing in T-odd channels is poorly
understood.  No existing codes can do more than HFODD in odd-$A$
octupole-deformed nuclei.

As above, we use the Skyrme interactions SIII, SkM$^*$, SLy4, and SkO'.  The
reason SkO' is our favorite has to do with the part of the energy functional
composed of T-odd spin densities (which, following common practice, we refer
to as the ``T-odd functional", even though the entire functional must be even
under T).  The T-odd functional plays no role in the mean-field ground states
of even nuclei, but can be important in any state with nonzero angular
momentum.  Of the forces above, only SkO' has been seriously investigated in
T-odd channels.  In Ref.\ \cite{bender02}, the T-odd part of the functional
was adjusted to reproduce Gamow-Teller resonances, resulting in an effective
Landau parameter $g_0' = 1.2$.  In the isoscalar channel, the force was
adjusted to reproduce the commonly used value $g_0=0.4$ \cite{osterfeld92}.
Although there are not enough data to constrain other relevant parameters in
the functional, and although a very recent calculation starting from a
realistic nucleon-nucleon interaction \cite{zuo03}, while confirming the 
value
$g_0'=1.2$, finds $g_0=0.85$, SkO' is clearly the best available Skyrme
interaction for describing spin-spin interactions.  The corresponding T-odd
terms in the functional are precisely those that will polarize the spin in 
the
core.  There are other terms that could be added to the standard Skyrme
interaction and do the same thing ---- the tensor force for example --- but
they are almost never used and their effects still need to be investigated.

The parameters of the other three forces SIII, SkM$^*$, and SLy4 were 
adjusted
entirely to ground-state properties in even-even nuclei, and so the Landau
parameters $g_0$ and $g_0'$ were not fit.  Here we set them to zero by
treating the T-odd and T-even terms in the Skyrme functional independently, 
as
described in Ref.\ \cite{bender02}.  Only orbital terms, which are fixed by
gauge invariance \cite{dobaczewski95a} (a generalization of Galilean 
invariance), appear in the T-odd parts of these forces.

We rely on the crudest forms of projection.  For parity, that means none at
all, and for angular momentum it means inserting the rigid-rotor factor
$J/(J+1) = 1/3 $ in front of the intrinsic Schiff moment, as described above.
In other words, we use Eq.\ (\ref{eq:intr}), with the intrinsic state taken 
to
be the Hartree-Fock ground state produced by HFODD.  Just as in the
particle-rotor model, the intrinsic Schiff moment is given by the classical
expression Eq.\ (\ref{eq:mom}), but with $\rho_{\rm ch}$ equal to $e$ times
the Hartree-Fock ground-state proton density.  As already mentioned, the
Hartree-Fock approximation allows us, by summing over occupied orbits, to
write the intrinsic matrix element of the two-body potential $\hat{V}_{PT}$ 
as
the expectation value of an effective one-body operator $\hat{U}_{PT}$.
Because we now have a microscopic version of the ``core", 
this effective potential is more complicated than in Eq.\ (\ref{eq:sppot}), 
and it now acts on all the nucleons:
\begin{widetext}
\begin{equation}
\label{eq:finite} \hat{U}_{PT}=  {\displaystyle \frac{g}{2 m_{\pi}^2 m_N}} 
\sum_{i=1}^A \svec{\sigma}_i \tau_{z,i} 
\cdot
\svec{\nabla} \int {\rm d}^3r' \left( \frac{m_{\pi}^2 e^{-m_{\pi}
|\svec{r}-\svec{r}'|}}{4\pi |\svec{r}-\svec{r}'|} \right) \Big[ ( 
\bar{g}_0 
+ 2
\bar{g}_2 )
\rho_1(\svec{r}') - \bar{g}_1 \rho_0(\svec{r}') \Big] + \textrm{exch.}
\end{equation}
\end{widetext}
Here $\rho_0(\svec{r}) \equiv \rho_n(\svec{r})+ \rho_p(\svec{r})$ and
$\rho_1(\svec{r})\equiv \rho_n(\svec{r})- \rho_p(\svec{r})$ are the isoscalar 
and isovector densities.  The piece coming from exchange terms in the original
two-body interaction $\hat{V}_{PT}$ is nonlocal, just as in the usual
Hartree-Fock mean field, and we have not written it explicitly here (though 
we do below).

The code HFODD at present cannot evaluate the expectation value of a folded
potential like that above, which is due to the finite pion range.
Nevertheless, even in the zero-range approximation we can avoid several of
the assumptions --- proportionality of neutron and proton densities,
negligibility of exchange terms, and absence of core spin --- leading to
the extremely simplified potential in Eq.\ (\ref{eq:sppot}). The zero-range
approximation is equivalent to assuming the pion is very heavy, so that the
term involving the pion mass in Eq.\ (\ref{eq:finite}) becomes a delta
function.  Under that assumption, but none others, the exchange terms
become local and $\hat{U}_{PT}$ takes the form:
\begin{widetext}
\begin{eqnarray}
\label{eq:sppottrue} \hat{U}_{PT}(\svec{r}) \longrightarrow &-{\displaystyle 
\frac{g}{2 m_{\pi}^2 m_N}} & \Big\{ \sum_{i=1}^A\
\svec{\sigma}_i \tau_{z,i} \cdot \Big[ ( \bar{g}_0 + 2 \bar{g}_2 )
\svec{\nabla} \rho_1(\svec{r}) -  \bar{g}_1 \svec{\nabla} \rho_0(\svec{r})
\Big]
\\ \nonumber
&&+{\displaystyle \frac{1}{2}} \sum_{i=1}^A \svec{\sigma}_i \cdot \Big[ (-3 
\bar{g}_0 + \bar{g}_1
\tau_{z,i}) \svec{J}_0 (\svec{r}) + ( \bar{g}_1 +\bar{g}_0 \tau_{z,i} - 4
\bar{g}_2 \tau_{z,i} ) \svec{J}_1(\svec{r}) \Big] \Big\} ~.
\end{eqnarray}
\end{widetext}
Here $\svec{J}(\svec{r})$ is the ``spin-orbit" current, defined, e.g., in
Ref.\ \cite{bender02} and references therein, and the subscripts 0 and 1 
refer
to isoscalar and isovector combinations as they do for the density.  The 
terms
in $\hat{U}_{PT}$ that contain $\svec{J}$ are the exchange terms omitted
above.  We will evaluate them, but argue later that their effects are 
probably
small when the finite range is restored.  The terms containing the density
$\rho$ all come from the direct part of $\hat{V}_{PT}$.  We do not simplify
things further to obtain something like Eq.\ (\ref{eq:sppot}) because
$\rho_p$ is not really proportional to $\rho_n$ and the core nucleons do 
carry
some spin.  We will manage nevertheless, to compare our results with those of
Ref.\ \cite{spevak97}.  We will also estimate the effect of a finite pion
range on the direct terms, though our inability to do so more precisely at
present is the most significant shortcoming of this work.

HFODD works by diagonalizing the interaction in the eigenbasis of an
optimal anisotropic three-dimensional harmonic oscillator. For
$^{225}$Ra, algorithms developed in Ref.\ \cite{dobaczewski97a} give
oscillator frequencies of $\hbar\omega_z$=7.0625 and
$\hbar\omega_\perp$=8.6765~MeV in the directions parallel and
perpendicular to the elongation axis. The matrix
element of $\hat{U}_{PT}$ converges only slowly as we increase the
number of levels in the basis.  When the interaction polarizes the
core, it takes 2500 or more single-particle basis states to get
convergence. The basis then contains up to $N_z$=26
and $N_\perp$=21 oscillator quanta.  
%
%
\subsection{Laboratory Schiff Moment of $^{225}$Ra}
We turn finally to results in $^{225}$Ra itself. For SkO', our HFODD
calculations yield $\beta_2=0.190$, $\beta_3=0.146$, and
$\beta_4=0.136$ for the usual first order approximation to the
deformation parameters determined from mass multipole moments
\cite{ring80}. The laboratory Schiff moment, Eq.\ (\ref{eq:intr}), is
proportional to the product of the intrinsic Schiff moment $\langle
\hat{S}_z \rangle$ and the expectation value $\langle \hat{V}_{PT}
\rangle$.  Table \ref{tab:1} shows the intrinsic moments and the
expectation values of the 6 operators that enter the zero-range
approximation to $\hat{V}_{PT}$ in Eq.\ (\ref{eq:sppottrue}). Before
commenting on the entries, we mention what is in each of the forces
and calculations.

For all the forces, terms in the functional that are proportional to
Laplacians of spin densities ($\svec{s}\cdot\Delta\svec{s}$) and
density-dependent spin-spin terms ($f(\rho)\svec{s}\cdot\svec{s}$), cf.\ 
Ref.\
\cite{dobaczewski95a,bender02}, which enter through the T-odd part of the
Skyrme functional, have been turned off.  For the first three lines in Table
\ref{tab:1} [forces labeled with (0)], the spin-spin terms have also been
turned off, so that the only nonzero terms in the T-odd functional (as noted
above) are those required by gauge invariance.  For the fourth line 
[SkO'(---)], all T-odd terms in the functional have been turned
off, so that aside from the self-consistency in the wave functions the
calculation resembles one with a phenomenological (non-self-consistent)
potential, for which T-odd mean-fields are never considered.  We include this
result so that we can distinguish the role played by core polarization.  The
results in the line labeled SkO' include the time-odd channels, adjusted as
discussed above \cite{bender02}.  This is the force in which we have the most 
confidence.  The
last entry is the result of Ref.\ \cite{spevak97}, with the implicit
assumption that the neutron and proton densities are proportional.

\begin{table*}[htp]
\caption{\label{tab:1}The intrinsic Schiff moment, in units of e\,fm$^3$
and the intrinisic-state expectation values of operators in
Eq.\ (\ref{eq:sppottrue}), in units of $10^{-3} \textrm{fm}^{-4}$.}
\begin{ruledtabular}
\begin{tabular}{lc|cc|cccc}
 & $\langle \hat{S}_z \rangle$
 & $\langle \svec{\sigma} \tau \cdot \svec{\nabla}\rho_0 \rangle$
 & $\langle \svec{\sigma} \tau \cdot \svec{\nabla}\rho_1 \rangle$
 & $\langle \svec{\sigma}      \cdot \svec{J}_0          \rangle$
 & $\langle \svec{\sigma}      \cdot \svec{J}_1          \rangle$
 & $\langle \svec{\sigma} \tau \cdot \svec{J}_0          \rangle$
 & $\langle \svec{\sigma} \tau \cdot \svec{J}_1          \rangle$ \\
\hline
SIII(0)   & 34.6& $-$1.081& $-$0.087& $-$1.047&    0.162& $-$1.049&    0.159 
\\
SkM$^*$(0)& 46.6& $-$0.730& $-$0.497& $-$1.043&    0.099& $-$1.042&    0.085 
\\
SLy4(0)   & 43.4& $-$0.676& $-$0.578& $-$1.303& $-$0.016& $-$1.299& $-$0.019 
\\
SkO'(---) & 41.7& $-$0.343& $-$0.318& $-$1.149&    0.030& $-$1.149&    0.030 
\\
\hline
SkO'   & 41.7& $-$0.467& $-$0.227& $-$0.476&    0.014& $-$0.526&    0.014 
\\
\hline
Ref.\ \protect\cite{spevak97}
         & 24   & $-$2    & $-$0.4  &      ---&      ---&      ---&      --- 
\\
\end{tabular}
\end{ruledtabular}
\end{table*}

In our calculations, the intrinsic Schiff moments are close to one another,
and all are less than twice the estimate of Ref.\ \cite{spevak97}.  The
agreement reflects the collective nature of these intrinsic moments; they are
even larger than the particle-rotor estimates.  But the
matrix elements of $\hat{V}_{PT}$, the other ingredient in Eq.\
(\ref{eq:intr}) for the laboratory Schiff moment, are a bit more delicate.
Our results show the exchange terms on the right side of the table to be
comparable to the direct terms, a result that is surprising because for a
spin-saturated core (or in the particle-rotor model) the exchange terms 
vanish
exactly.  We think, however, that the ratio of exchange to
direct terms would become small were the finite range of the interaction 
reintroduced and short-range NN correlations inserted.

Though unable to include either effect here, we did so in a Nilsson model for
$^{225}$Ra.  We took nucleons there to occupy independent single-particle
levels generated by a deformed oscillator potential with $\beta_2=0.138$,
$\beta_3=0.104$, and $\beta_4=0.078$, values taken from Ref.\ \cite{spevak97}.
We then evaluated the ground-state expectation value of the full two-body
interaction $\hat{V}_{PT}$, with and without the zero-range approximation (and
in the latter case, with short-range correlations included {\it \`a la} Ref.\
\cite{miller76}).  In this simple model, the valence nucleon carries all the
spin, and only the neutron-proton and neutron-neutron parts of $\hat{V}_{PT}$
contribute.  The direct $np$ term shrank by a factor of 1.5, while the
corresponding exchange term shrank by a factor of 1400 (both independently of
the $\bar{g}$'s in Eq.\ (\ref{eq:pion}), it turns out) when the range of the
interaction was set to its proper value.  The results in the $nn$
channel were less dramatic:  the direct part again shrank by 1.5 and the
exchange part by a factor of 5.  When we moved the valence neutron to higher
orbits, these numbers changed some --- the direct terms sometimes were not
suppressed at all and other times shrank by factors of up to 6, but the ratios
of the exchange to direct contributions almost always ended up small.  Similar
behavior was found for parity-violating forces in Ref.\ \cite{adelberger85},
where it was traced in part to the different average momenta carried by the
pion in direct and exchange graphs.  So that we can compare our results with
those of Ref.\ \cite{spevak97}, we will neglect the exchange terms from now
on, though we caution that this step should eventually be justified more
rigorously, e.g., by actually calculating them with the finite-range force
in the full mean-field model.
The reduction we see in the direct terms is in line with the results of Ref.\
\cite{griffiths91}, though we find it more variable\footnote{We performed
another test, using the direct part of Eq.\ (\ref{eq:sppottrue}) with the
valence wave function taken from the Nilsson model just described, but with
the neutron and protons densities assumed to have more realistic Wood-Saxon
forms.  The direct terms were again suppressed by factors of 1.5 to almost 10
that depended significantly on the valence orbit.}.

Though we cannot yet be more quantitative about finite-range effects, we do
quantify the core polarization in Table \ref{tab:1}.  For the first three
lines of the table, where the forces are labeled (0), the spin-spin terms are
absent from the energy functional, and the protons in the core develop only a
tiny spin density from the T-odd terms required by gauge invariance.  For
the fourth line, SkO'(---), all T-odd terms are absent and the protons can
have no spin at all.  This means that the operators $f(\svec{r})
\svec{\sigma}$ and $f(\svec{r}) \svec{\sigma} \tau$ have either the same or
almost the same expectation value for any $f(\svec{r})$ so that columns 4 and
6 ($\langle \svec{\sigma} \cdot \svec{J}_0 \rangle$ and $\langle \svec{\sigma}
\tau \cdot \svec{J}_0 \rangle$) have identical or nearly identical entries for
these forces, and so do columns 5 and 7 ($\langle \svec{\sigma} \cdot
\svec{J}_1 \rangle$ and $\langle \svec{\sigma} \tau \cdot \svec{J}_1
\rangle$).  The fifth line of the table contains the effects of spin
polarization, which are primarily to alter the neutron-spin density; the
equalities between the columns are not badly broken, so the protons do not
develop much spin.  The same is true of the terms involving $\rho$, though
that is not obvious from the table because we display only the two terms that
appear in Eq.\ (\ref{eq:sppottrue}).

These near equalities and the probable irrelevance of the exchange terms when
the finite range is taken into account imply that only the quantities
$\svec{\sigma}_n \cdot\bm{\nabla} \rho_n$ and $\svec{\sigma}_n
\cdot\bm{\nabla} \rho_p$ are ultimately important.  We display them in Table
\ref{tab:2}.  Except for SIII, the neutron-density distribution affects the
matrix element much more than the that of protons.  By comparing the fourth
and fifth lines, however, we see that spin correlations increase the role
of the protons, while reducing that of the neutrons slightly.  Thus, while the
spin-spin interactions do not cause the protons to develop much net spin, they
do correlate the neutron spin with the proton density.

\begin{table}[htp]
\caption{\label{tab:2} Intrinsic-state expectation values of important matrix
elements, in the neutron-proton scheme, in units of $10^{-3} 
\textrm{fm}^{-4}$.}
\begin{ruledtabular}
\begin{tabular}{lcc}
 & $\langle \svec{\sigma}_n \cdot \svec{\nabla}\rho_n \rangle$
 & $\langle \svec{\sigma}_n \cdot \svec{\nabla}\rho_p \rangle$ \\
\hline
SIII(0)   &  $-$0.577 &  $-$0.491 \\
SkM$^*$(0)&  $-$0.619 &  $-$0.120 \\
SLy4(0)   &  $-$0.628 &  $-$0.050 \\
SkO'(---) &  $-$0.331 &  $-$0.013 \\
\hline
SkO'   &  $-$0.320 &  $-$0.114 \\
\hline
Ref.\ \protect\cite{spevak97}
         &   $-$1.2   &  $-$0.8   \\
\end{tabular}
\end{ruledtabular}
\end{table}

There is not too much scatter in our results.  The entries in the second 
column ($\langle \svec{\sigma} \tau \cdot \svec{\nabla}\rho_0 
\rangle$) of Table \ref{tab:1} differ by factors of two or three, and the
entries in the third ($\langle \svec{\sigma} \tau \cdot \svec{\nabla}\rho_1 
\rangle$) by a little more, though they are all smaller
than those in the second column (which is not surprising --- the third column 
subtracts the neutron and proton densities while the second adds
them).  In the neutron-proton scheme (table \ref{tab:2}) all of our numbers
are smaller than those of Ref.\ \cite{spevak97}, a result that was
anticipated in Ref.\ \cite{engel99a}.  The difference from the earlier
estimate for the larger matrix elements ranges from factors of two to four,
though the isovector combination --- the third column in table \ref{tab:1}
--- is sometimes actually enhanced a little.  

What, at last, have we to say about the real laboratory Schiff moment $S$?
The lab moment is given by the product of the matrix elements just discussed,
the intrinsic Schiff moments, and the unknown coefficients $\bar{g}_i$.  Our
intrinsic Schiff moments are about 1.5 times larger than those of Ref.\
\cite{spevak97}, while our $\hat{V}_{PT}$ matrix elements, in the zero-range
approximation, are smaller than theirs, usually by a somewhat larger amount.
Overall, our lab moments will usually be smaller by factors between
about 1.5 and 3 than the estimates of Ref.\ \cite{spevak97} (an exception can
occur if for some reason $\bar{g}_1$ is considerably less than the other two
coefficients).

How big are our moments compared to that of $^{199}$Hg?  The most 
comprehensive calculation in that nucleus, which appeared very recently 
\cite{dmitriev03}, improved on the work of Ref.\ \cite{flambaum86} by 
including the effects of the residual strong interaction and the full 
finite-range form for $\hat{V}_{PT}$.  The new results are smaller than that
of ref.\ \cite{flambaum86}, only slightly so for the isovector part 
of $\hat{V}_{PT}$, but by a considerably amount in the isoscalar and isotensor
channels.
The authors write their results in terms of the 
pion-nucleon couplings as
\begin{equation}
\label{eq:SHg}
S_{\rm Hg} = .0004 \, g \bar{g}_0  +.055 \, g  \bar{g}_1 +.009 \, g \bar{g}_2 
~[e \, {\rm 
fm}^3].
\end{equation}
Our result for radium, with the zero-range approximation and exchange terms
neglected, translates to
\begin{equation}
\label{eq:SRa}
S_{\rm Ra}^{\rm zero-range} = -5.06 \, g \bar{g}_0 +10.4 \, g  \bar{g}_1 -10.1 
\, g \bar{g}_2 ~[e \, {\rm fm}^3].
\end{equation}
If the three $\bar{g}$'s are comparable, therefore, our Schiff moment is 
several hundred times larger than that of Ref.\ \cite{dmitriev03}, in part
because the isoscalar and isotensor interactions are more effective in
Ra than in Hg. [If $\bar{g}_1$ is larger than the other two
couplings, as in left-right symmetric models \cite{herczeg87}, our result is 
less than 200 times bigger than 
the latest one in $^{199}$Hg. The very small coefficient of $\bar{g}_0$ for 
$^{199}$Hg in Eq.\ (\ref{eq:SHg}), by the way, has significant consequences 
\cite{herczeg87} for the limit on the QCD T-violating parameter $\bar{\theta}$ 
that can be inferred from the experimental limit in Ref.\ \cite{romalis01}.]  
Accepting 
the work of Ref.\ \cite{dzuba02} on atomic physics in Ra and Hg,
the enhancement of the atomic EDM of $^{225}$Ra is about about three times 
that of the Schiff moment, i.e.\ potentially more than 1000.  We again 
caution, however, that we have 
yet to include the full 
finite-range version of $\hat{V}_{PT}$ and that our preliminary investigations 
suggest that doing so will decrease our Schiff moment at least a little.  
Ironically, Ref.\ \cite{dmitriev03} finds that including the finite range 
actually increases the matrix element in $^{199}$Hg, though only slightly.  

We hope to make other improvements in our calculation as well.  Projection 
onto states of good parity will change the results a bit, as will 
angular-momentum projection. Our
conclusions about the size of spin-polarization corrections could be
modified by two terms in the Skyrme functional we've set to zero, or by a 
better determined value of the Landau parameter $g_0$.  Whatever the result of
such corrections, however, it is clear that the atomic EDM of $^{225}$Ra will
always be significanlty larger than that of $^{199}$Hg.
%
\section{\label{s:5}Conclusions}
We have calulated the Schiff moment in $^{225}$Ra in the approximation that 
the T-violating interaction has zero range.  Our calculations, which are 
self-consistent and include core polarization, give results that are generally 
just a few times smaller than earlier estimates based on the particle-rotor
model.  
Accepting the very recent results of Ref.\ \cite{dmitriev03}, we currently 
find the Schiff 
moment of $^{225}$Ra to be (generically) several hundred times that of 
$^{199}$Hg, a result that strengthens the case for an atomic-EDM experiment in
Ra, though the enhancement factor depends significantly on the source 
of T violation, and we 
expect it to decrease at least a little when we use the finite-range 
force.  Work towards including a finite range in HFODD is in progress.  We 
also plan to apply the self-consistent methods used here to other light 
actinides, as well as to $^{199}$Hg, where we suspect octupole correlations 
may play some role \cite{engel99a}.  Maintaining self consistency in 
$^{199}$Hg should automatically control the spurious Schiff strength 
encountered in Ref.\ \cite{dmitriev03}.   The source of the insensitivity of
the
Schiff moment to T violation in the isoscalar channel in that work should
be checked and understood.

After many years of neglect, 
the question of which isotopes 
are best for EDM measurements is now being rapidly addressed.
%
%
\begin{acknowledgments}
This work was supported in part by the U.S.\ Department of Energy under
grant DE-FG02-02ER41216, by the Polish Committee for Scientific Research
(KBN) under Contract No.~5~P03B~014~21, and by computational
grants from the \emph{Regionales Hochschulrechenzentrum
Kaiserslautern} (RHRK) Germany and from the Interdisciplinary Centre
for Mathematical and Computational Modeling (ICM) of the Warsaw
University.  
M.~B.\ acknowledges support through a European Community Marie Curie
Fellowship, and J.~H.~J.\ acknowledges partial support through a Doctoral 
Fellowship from the Portuguese Foundation for Science and Technology.
\end{acknowledgments}

\end{document}